\documentstyle[pra,aps,twocolumn,psfig,floats,amssymb]{revtex}

\def\be{ \begin{equation} }
\def\ee{ \end{equation} }
\def\ba{ \begin{array} }
\def\ea{ \end{array} }
\def\bea{ \begin{eqnarray} }
\def\eea{ \end{eqnarray} }
\def\bml{ \begin{mathletters} }
\def\eml{ \end{mathletters} }
\def\bmla{ \bml \bea }
\def\emla{ \eea \eml }

\def\eps{\varepsilon}
\def\Tdj{\tau_{\rm d}^{\rm jump}}
\def\Tdr{\tau_{\rm d}^{\rm relax}}
\def\Taj{\tau_{\rm a}^{\rm jump}}
\def\Tar{\tau_{\rm a}^{\rm relax}}
\def\Taji{\tau_{\rm a}^{\rm jump, i}}
\def\Tajf{\tau_{\rm a}^{\rm jump, f}}
\def\Wd{w_{\rm d}}
\def\Wa{w_{\rm a}}
\def\Wat{W_{\rm a}}
\def\thi{\vartheta}
\def\th{\thi(\tau)}
\def\hi{\chi}
\def\hio{\hi(\omega)}
\def\Pd{P_{\rm d}}
\def\Pdt{\Pd(\tau)}
\def\Pdi{\Pd(\infty)}
\def\Pa{P_{\rm a}}
\def\Pat{\Pa(\tau)}
\def\Pai{\Pa(\infty)}

\begin{document}
\draft
\wideabs{
\author{N. V. Vitanov \cite{email}}
\address{Helsinki Institute of Physics, P.O. Box 9,
 FIN-00014 University of Helsinki, Finland}
\title{Transition times in the Landau-Zener model}
\date{\today }
\maketitle
\begin{abstract}
This paper presents analytic formulas for various transition times
in the Landau-Zener model.
Considerable differences are found between the transition times in the
diabatic and adiabatic bases, and between the jump time
(the time for which
the transition probability rises to the region of its asymptotic value)
and the relaxation time (the characteristic damping time of the oscillations
which appear in the transition probability after the crossing).
These transition times have been calculated by using the exact values of
the transition probabilities and their derivatives at the crossing point
and approximations to the time evolutions of the transition probabilities
in the diabatic basis, derived earlier
\protect{[}N. V. Vitanov and B. M. Garraway,
Phys. Rev. A {\bf 53}, 4288 (1996)\protect{]},
and similar results in the adiabatic basis, derived in the present paper.
\end{abstract}
}

\section{Introduction}

\label{Sec-introduction}

The Landau-Zener (LZ) model \cite{Landau32,Zener32} has long ago become
a standard notion in quantum physics.
It provides the probability of transition between two quantum states coupled
by an external field of a constant amplitude and a time-dependent frequency
which passes through resonance with the transition frequency.
This {\it level crossing}, seen in the diabatic basis (the basis of the two
bare states -- the eigenstates of the Hamiltonian in the absence of
interaction), appears as an {\it avoided crossing} in the adiabatic basis
(the basis comprising the two eigenstates of the Hamiltonian in the presence
of interaction).
Cases of level crossings and avoided crossings can be met in a number of
areas in physics, such as
quantum optics,
magnetic resonance,
atomic collisions,
solid-state physics,
atom-surface scattering,
molecular physics, and
nuclear physics.
The Landau-Zener model is a reliable qualitative (and often even quantative)
tool for describing and understanding such phenomena.

Along with the transition probability, it is often necessary to know the
time for which the transition occurs -- the {\it transition time}.
For example, in the case of multiple level crossings (or avoided crossings) it
is essential to know if the transition is completed before the next crossing.
Moreover, even in the case of a single crossing the transition time is
an important parameter because the actual coupling and detuning never match
the LZ ones exactly.
For instance, the LZ formula applies when the transitions take place
in a narrow time interval around the crossing, provided the actual detuning
changes nearly linearly with time and the coupling is nearly constant
in the vicinity of the crossing.
However, when transitions can occur far from the crossing, where the actual
coupling usually is not constant and the detuning is not linear
in time, the usage of the LZ formula may be incorrect.

It is far from obvious what is the transition time $\tau_d$ in the
{\it diabatic} basis because there the coupling $\Omega$ is constant
and lasts from $-\infty$ to $+\infty$.
Since the detuning, being a linear function of time, $\Delta=\beta^2t$,
diverges at $\pm\infty$, the transition probability is well defined,
but there is no apparent time scale at which the transition takes place.
It looks considerably easier to determine the transition time $\tau_a$ in the
{\it adiabatic} basis because there the (non-adiabatic) coupling is a
Lorentzian function of time with a width of $2\Omega/\beta^2$ and the
energies of the two adiabatic states have an avoided crossing
with the same duration;
hence, the transition time is expected to be $\tau_a\approx 2\Omega/\beta^2$.
This deceptively obvious conclusion turns out to be only partially correct.
The problem here arises from the fact that the non-adiabatic coupling
vanishes too slowly and also, the eigenenergy gap diverges too slowly.

The {\it scaling} properties of the LZ transition time $\tau_d$ in the
{\it diabatic} basis have been studied by Mullen {\it et al} \cite{Mullen89},
who have found that for large $\Omega$, $\tau_d$ is proportional
to $\Omega $, while for small $\Omega $, $\tau_d$ is constant.
These authors have used two different approaches -- an ``internal clock'',
based on perturbative calculation of the transition probability time evolution,
and an ``external clock'', based on identifying a characteristic
frequency appearing in the response of the two-state system to a harmonic
perturbation.

In the present paper, I derive analytical estimates for the LZ transition
times by using some exact and approximate results for the transition
probability in the diabatic basis, derived in \cite{Vitanov97}, and
similar results in the adiabatic basis derived here.
Thus, this paper provides not only the scaling properties but also explicit
formulas for the transition times in both the diabatic and adiabatic bases.
In view of the numerous exact and approximate results for the transition
probabilities, it turns out much easier to calculate the transition time
than to define it.
I distinguish two kinds of transition times.
The {\it jump time} $\tau^{\rm jump}$ is the
time for which the transition probability rises to the region of its
asymptotic value $P(\infty )$ (the exact definition is given below).
The  {\it relaxation time} $\tau^{\rm relax}$ is the time for which the
amplitude of the oscillations, which appear in the transition probability
after the crossing, is damped to a sufficiently small value,
$\eps P(\infty),\ (\eps \ll 1)$.

The paper is organized as follows.
The basic equations and definitions are given in Sec. \ref{Sec-definition}.
The transition times in the diabatic basis are calculated
in Sec. \ref{Sec-diabatic}
and those in the adiabatic basis in Sec. \ref{Sec-adiabatic}.
The conclusions are summarized in Sec. \ref{Sec-conclusion}.

\section{Basic equations and definitions}

\label{Sec-definition}

\subsection{Diabatic basis}

The time evolution of a coherently driven two-state quantum system is
described by the Schr\"odinger equation (in units $\hbar =1$) \cite{Shore90}
\be
\label{SEq}
i\frac d{dt} \left[ \ba{c} c_1(t) \\ c_2(t) \ea \right] =
	\left[ \ba{cc} -\Delta (t) & \Omega (t) \\ 
			\Omega (t) & \Delta (t) \ea \right] 
	\left[ \ba{c} c_1(t) \\ c_2(t) \ea \right],
\ee
where $c_1(t)$ and $c_2(t)$ are the probability amplitudes of states
$\psi _1$ and $\psi _2$, $\Omega (t)$ is the coupling (assumed real) between
the two states and $\Delta (t)$ is a half of the difference between the system
transition frequency and the field frequency.
In the Landau-Zener model, we have
\be
\label{LZ}
\Omega (t)={\rm const},\qquad \Delta (t)=\beta ^2t,
\ee
where the coupling $\Omega (t)$ is supposed to last
from $t\rightarrow -\infty $ to $t\rightarrow +\infty $.
Without loss of generality the slope $\beta ^2$ of the detuning
as well as the real constants $\Omega$ and $\beta$ are assumed positive.
Both $\Omega$ and $\beta$ have the dimension of frequency.
Following the notation of \cite{Vitanov97}, I introduce the scaled
dimensionless time $\tau$ and coupling $\omega$, 
\be
\label{tau-omega}
\tau=\beta t,\qquad \omega =\frac \Omega \beta .
\ee
Provided the system has been initially in state $\psi _1$, the probability
of transition to state $\psi _2$ at time $\tau$, 
$\Pdt=\left|c_2(\tau)\right| ^2$, is \cite{Vitanov97,Shore90,Lim91} 
\be
\label{Pd-exact}
\Pdt = \frac{\omega ^2}2e^{-\pi \omega ^2/4}
\left| D_{-1+i\omega ^2/2}\left(\tau\sqrt{2}e^{3i\pi /4}\right) \right| ^2,
\ee
where $D_\nu (z)$ is the parabolic cylinder function \cite{Erdelyi,AS}.
Here the subscript ``$d$'' indicates the {\it diabatic} basis.

\subsection{Adiabatic basis}

The two adiabatic states (defined as the instantaneous eigenstates of
the Hamiltonian) are given by
$\varphi_1(\tau)=\psi_1\cos\th-\psi_2\sin\th$,
$\varphi_2(\tau)=\psi_1\sin\th+\psi_2\cos\th$,
where the angle $\th$ is defined as 
\be
\label{theta}
\tan 2\th =\frac{\Omega (\tau)}{\Delta (\tau)}
	  =\frac \omega \tau,
	\qquad (0\leqq \th \leqq \pi /2).
\ee
The eigenvalues of states $\varphi_1$ and $\varphi_2$ are given by
$-\Omega_0$ and $+\Omega_0$, respectively, and the (non-adiabatic)
coupling between them by $\thi^\prime(\tau) \equiv d\th/d\tau$, where
\be
\label{Omega0-dtheta}
\Omega _0(t)=\sqrt{\tau^2+\omega ^2},\qquad
\thi^\prime(\tau)=-\frac \omega {2(\tau^2+\omega ^2)}. 
\ee
The condition for adiabatic evolution is
$\label{AdiabCond}\left| \thi^\prime\right| \ll \Omega _0$,
which requires that $\omega^2 \gg 1$.
Hence, the coupling $\omega \equiv \Omega/\beta$ plays also the role
of the adiabaticity parameter.

The transition probability $\Pat$ between the two adiabatic states
can be obtained from the relation between the evolution matrices
in the adiabatic and diabatic bases by using the exact diabatic
evolution matrix given in \cite{Vitanov97}
and some properties of $D_\nu(z)$ \cite{Erdelyi}.
The result is
\bea
\label{Pa-exact}
&&\Pat = e^{-\pi \omega ^2/4}
	\bigg| D_{i\omega ^2/2}\left(\tau\sqrt{2}e^{3i\pi /4}\right)
			 \cos \th \nonumber \\
&&\qquad - \frac \omega {\sqrt{2}} e^{-i\pi /4}
		D_{-1+i\omega ^2/2}\left( \tau\sqrt{2}e^{3i\pi /4}\right)
			\sin \th \bigg| ^2.
\eea

In the figures below, both the diabatic and adiabatic transition probabilities
are calculated by the highly accurate numerical method desribed in the 
Appendix, rather than from Eqs.\ (\ref{Pd-exact}) and (\ref{Pa-exact}).
These equations are used for analytic derivation of the transition times.

\subsection{Some exact values of the transition probabilities}

For the calculation of the transition times, we need a few values
of $\Pdt$ and $\Pat$, easily obtained from Eqs.\ (\ref{Pd-exact})
and (\ref{Pa-exact}).
By taking the limit $\tau\rightarrow \infty $, we recover the well-known
LZ probabilities
\be
\label{Pd(inf)}
\Pdi=1-e^{-\pi \omega ^2}, 
\ee
\be
\label{Pa(inf)}
\Pai=e^{-\pi \omega ^2}. 
\ee
By using the power series \cite{Vitanov97,AS,Abadir93} of $D_\nu(z)$
we can expand $\Pdt$ and $\Pat$ in terms of $\tau$, which enables us
to find the values of $\Pdt$ and  $\Pat$ and their derivatives at $\tau=0$.
We need the following values
\bml
\label{dPd}
\bea
\label{Pd(0)}
&&\Pd(0)=\case12\left( 1-e^{-\pi \omega ^2/2}\right) , \\
\label{dPd(0)}
&&\Pd^{\prime }(0)=\omega \sqrt{1-e^{-\pi \omega ^2}}\cos \hi , \\
\label{d2Pd(0)}
&&\Pd^{\prime \prime }(0)=2\omega ^2e^{-\pi \omega ^2/2},
\eea
\eml
and
\bml
\label{dPa}
\bea
\label{Pa(0)}
&&\Pa(0)=\case12\left( 1-\sqrt{1-e^{-\pi \omega ^2}}\sin \hi \right) , \\
\label{dPa(0)}
&&\Pa^{\prime }(0)=\frac 1{2\omega }e^{-\pi \omega ^2/2}, \\
\label{d2Pa(0)}
&&\Pa^{\prime \prime }(0)= \sqrt{1-e^{-\pi \omega ^2}}
	\left( \frac {\sin \hi}{2\omega^2} -\cos \hi \right), 
\eea
\eml
where the angle $\hio$ is defined by 
\be
\label{chi}
\hio = \frac \pi 4 
	+ \arg \Gamma \left( \case12-\case14i\omega^2\right)
	- \arg \Gamma \left( 1-\case14i\omega^2\right) .
\ee
It is a monotonically increasing function of $\omega $.
For small or large $\omega $, $\hio$ behaves as 
\bmla
\label{chi-small}
&&\hio = \frac \pi 4 + \frac{\ln 2}2\omega^2
	\!- \frac{\zeta(3)}{32}\omega^6\!
	+{\cal O}(\omega^{10}),\ (\omega^2\! \ll \!1),\\
\label{chi-large}
&&\hio = \frac \pi 2 - \frac 1{2\omega^2} - \frac 1{3\omega^6} 
	+{\cal O}(\omega^{-10}),\ (\omega^2 \gg 1),
\emla
where $\zeta(z)$ is the Riemann's zeta function \cite{AS}.

\section{Transition times in the diabatic basis}

\label{Sec-diabatic}

\subsection{Time evolution of the transition probability}

\begin{figure}[t]
\vspace*{0mm}
\centerline{\psfig{width=85mm,file=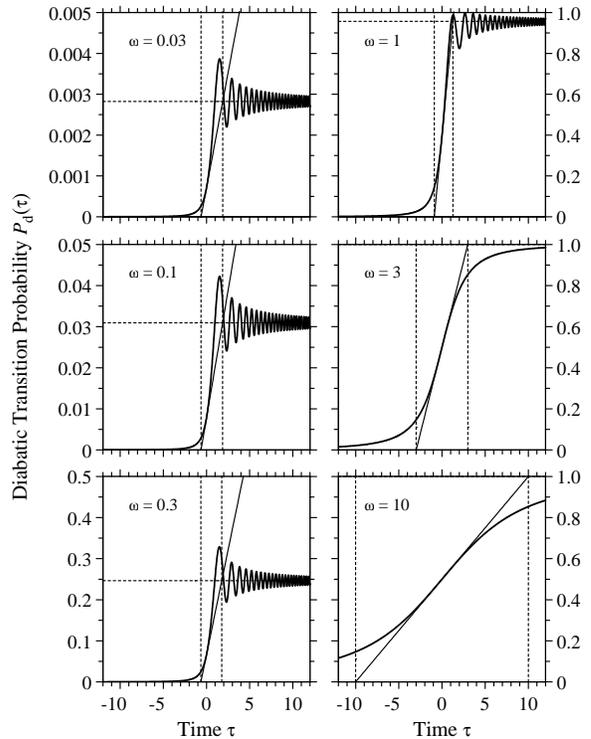}}
\vspace*{0mm}
\caption{
The time evolution of the diabatic transition probability $\Pdt$
(thick curve) for six values of the coupling $\omega$ -- from
$\omega=0.03$ (small) to $\omega=10$ (large).
In each figure, the horizontal dashed line shows the asymptotic value
$\Pdi$, Eq.\ (\protect\ref{Pd(inf)}) \protect{[}for $\omega=3$ and 10,
it coincides with the axis $\Pdt=1$\protect{]}.
The two vertical dashed lines display the jump region,
whose beginning is defined by the crossing of the tangent to $\Pdt$ at
$\tau=0$ (shown by a slanted solid line) with the zero line, whereas its
end is defined by the crossing of the tangent with the $\Pdi$ line.
}
\label{Fig-diabatic}
\end{figure}

The time evolution of the diabatic transition probability $\Pdt$ is
shown in Fig.\ \ref{Fig-diabatic} for six values of the coupling $\omega$
-- from 0.03 (small) to 10 (large).
The evolution shows two characteristic time regions.
The first one is around the crossing ($\tau=0$), where $\Pdt$ rises from zero
to about its asymptotic value $\Pdi$;
this region determines the {\it jump time} $\Tdj$.
This jump is followed by a region, where $\Pdt$ oscillates around
the value $\Pdi$ (for large $\omega$, these oscillations become invisible);
this region determines the {\it relaxation time} $\Tdr$.

The time evolution of $\Pdt$ before and after the crossing is well
approximated by the formulas \cite{Vitanov97}
\bml
\label{Pd}
\bea
\label{Pdm}
&&\Pd(\tau<0) \approx \case12 + \frac \tau{2 \sqrt{\tau^2+\omega ^2}},\\
\label{Pdp}
&&\Pd(\tau>0) \approx \case12 + \left( \case12 - e^{-\pi \omega^2}\right)
	\frac \tau{\sqrt{\tau^2+\omega ^2}} \nonumber\\
&&\qquad\qquad - e^{-\pi \omega ^2/2}\sqrt{1-e^{-\pi \omega ^2}}
	   \frac{\omega}{\sqrt{\tau^2+\omega ^2}} \cos \xi (\tau),
\emla
valid for $\tau^2+\omega^2\gtrsim1$. The phase $\xi(\tau)$ is given by
\bea
\label{xi}
&&\xi (\tau) = -\frac{\omega ^2}2
     +\omega ^2\ln \left[ \case 1{\sqrt{2}}
	\left( \tau+\sqrt{\tau^2+\omega ^2}\right)\right]\nonumber\\
&&\qquad\qquad +\tau\sqrt{\tau^2+\omega ^2}
     +\frac \pi 4
     +\arg \Gamma \left( 1-\case12i\omega ^2\right) .
\eea

\subsection{Jump time}

The attempt to define the jump time in a simple and unambiguous manner
quickly comes across some difficulties.
First of all, it is not obvious how to define the {\it initial} time of
the transition, because $\Pdt$ is nonzero for any finite time.
A reasonable choice seems to be the time
when the rising transition probability $\Pdt$ first equals $\eps \Pdi$,
where $\eps$ is a suitably chosen small positive number ($\eps \ll 1$). 
It is even less obvious how to define the {\it final} time of the transition
in a way that applies to both small and large coupling.
One possible choice, used by Lim and Berry \cite{Lim91}, is to take the time
at which the upper envelope of the oscillations [obtained by setting
$\cos\xi(\tau)=-1$ in Eq.\ (\ref{Pdp})] touches unity.
However, this idea does not apply for small $\omega$, because then the
upper envelope never touches unity.
Another possibility is to take the time at which $\Pdt$ crosses its
asymptotic value $\Pdi$ for the first time.
However, this is appropriate for small $\omega$ only, because for large
$\omega$ (when the oscillations are strongly damped),
the crossover takes place at exponentially large times,
although $\Pdt$ comes very close to $\Pdi$ much earlier.
Alternatively, one can take the time when $\Pdt$
first equals the value $(1-\eps) \Pdi$, which is reasonable and
consistent with the $\eps$-definition of the initial time discussed
above, but leads to a rather complicated expression.
I propose here a more elegant and simple solution, which applies to
any $\omega$ and provides similar results as the $\eps$-approach.
I define the jump time as
\be
\label{Td-jump-def}
\Tdj = \frac{\Pdi}{\Pd^{\prime}(0)}. 
\ee
This definition is based upon the geometrical meaning of the derivative
as the slope of the function at the point of calculation.
It provides good results when applied to the most frequently used smooth
functions that rise monotonically from 0 to 1,
e.g. $f(x)=\case12(1+{\rm tanh}x)$ and $f(x)=\case12(1+x/\sqrt{x^2+1})$,
typically providing the interval where the function rises from about
$0.10 - 0.15$ to about $0.85 - 0.90$.
By using the exact values of $\Pdi$ and $\Pd^{\prime }(0)$ from
Eqs.\ (\ref{Pd(inf)}) and (\ref{dPd(0)}), we obtain
\be
\label{Td-jump}
\Tdj = \frac{\sqrt{1-e^{-\pi \omega ^2}}}{\omega \cos \hio }. 
\ee
It follows from Eqs.\ (\ref{chi-small}) and (\ref{chi-large}) that
\bml
\bea
\label{Td-jump-small}
&&\Tdj \approx \sqrt{2\pi }
	,\qquad (\omega^2 \ll 1), \\
\label{Td-jump-large}
&&\Tdj \approx 2\omega
	,\qquad (\omega^2 \gg 1). 
\eea
\eml
In other words, the jump time is proportional to $\omega $ at large
$\omega $ while it is nearly constant for small $\omega $.
Thus, Eqs.\ (\ref{Td-jump-small}) and (\ref{Td-jump-large}) confirm the
scaling properties found in \cite{Mullen89} for the extreme cases of small
and large $\omega$.

This behavior of the jump time for small and large $\omega$ can be
explained from Eqs.\ (\ref{Pdm}) and (\ref{Pdp}), which provide
$\Pd(\tau<0)$ and $\Pd(\tau>0)$, and from the Taylor expansion of $\Pdt$
around $\tau=0$, obtained by using the derivatives (\ref{dPd}).
It can easily be shown that for large $\omega$, $\Pdt$ depends on
the ratio $\tau/\omega$ only which means that $\Tdj \propto \omega$.
For small $\omega$, the normalized transition probability
$\Pdt/\Pdi$ depends on $\tau$ only, which can indeed be seen
in Fig.\ \ref{Fig-diabatic} for $\omega=0.03$, 0.1 and 0.3;
hence, $\Tdj$ must not depend on $\omega$.

\begin{figure}[t]
\vspace*{0mm}
\centerline{\psfig{width=70mm,file=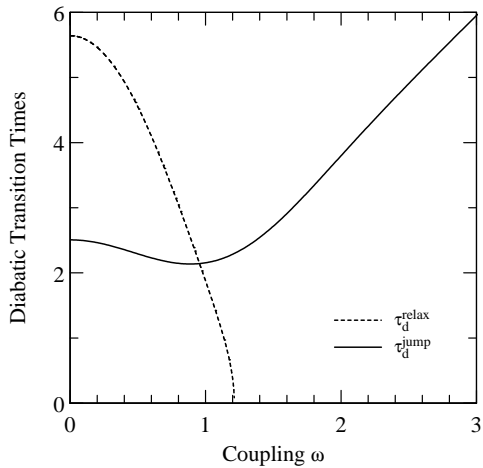}}
\vspace*{-1mm}
\caption{
The jump time (\protect\ref{Td-jump}) and the relaxation time
(\protect\ref{Td-relax}) (for $\eps=0.1$) of the diabatic transition
probability, plotted against the dimensionless coupling
$\omega \equiv \Omega/\beta$.
}
\label{Fig-Td}
\end{figure}

\subsection{Relaxation time}

As Eq.\ (\ref{Pdp}) shows, the amplitude of the oscillations in
$\Pd(\tau>0)$ vanishes as $\tau^{-1}$ at large positive times.
I define the relaxation time $\Tdr$ as the time it takes
to damp the oscillation amplitude to the (small) value $\eps \Pdi$,
where $\eps \ll 1$.
By using Eq.\ (\ref{Pdp}), we find
\be
\label{Td-relax}
\Tdr \approx \omega 
	\sqrt{\frac 1{\eps^2\left( e^{\pi \omega ^2}-1\right) }-1}.
\ee
The square root is real only for $\pi\omega^2 \leqq \ln(1/\eps^2 + 1)$.
This inequality imposes an upper limit of $\omega$, above which the
oscillation amplitude is never larger than $\eps \Pdi$.
For $\eps =0.1$, this limit is $\omega \lesssim 1.21$.

The diabatic jump time (\ref{Td-jump}) and the relaxation time
(\ref{Td-relax}) (with $\eps=0.1$) are displayed in Fig.\ \ref{Fig-Td}.

\begin{figure}[t]
\vspace*{0mm}
\centerline{\psfig{width=85mm,file=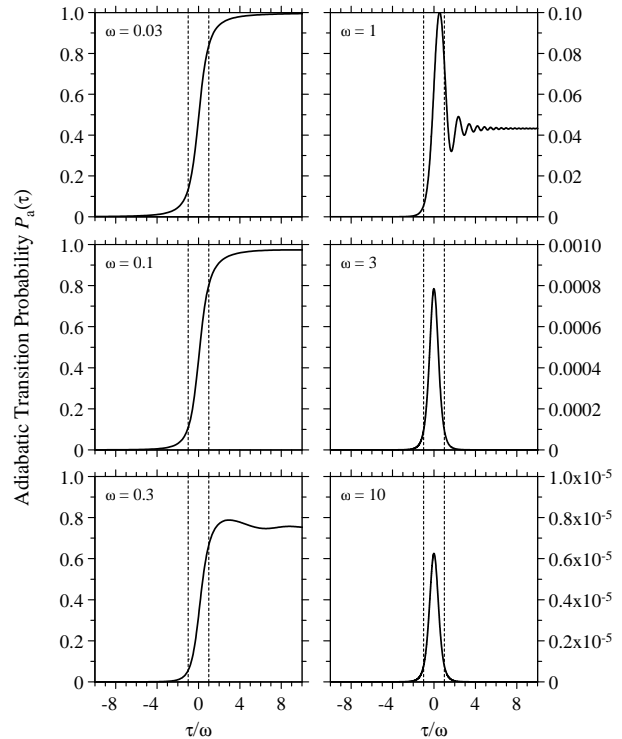}}
\vspace*{0mm}
\caption{
The adiabatic transition probability evolution $\Pat$
(solid curve) plotted against the scaled time $\tau/\omega$ for the same
six values of the coupling $\omega$ as in Fig.\ \protect\ref{Fig-diabatic} --
from $\omega=0.03$ (small) to $\omega=10$ (large).
The two vertical dashed lines in each figure display the interval
$-\omega \leqq \tau \leqq \omega$, which provides approximately
the jump region for $\omega \lesssim 1$.
}
\label{Fig-adiabatic}
\end{figure}

\section{Transition times in the adiabatic basis}

\label{Sec-adiabatic}

\subsection{Time evolution of the transition probability}

The time evolution of the adiabatic transition probability $\Pat$ is
shown in Fig.\ \ref{Fig-adiabatic} for the same six values of the coupling
$\omega$ as in Fig.\ \ref{Fig-diabatic} -- from 0.03 (small) to 10 (large).
There are two distinct types of evolution.
For small $\omega$, $\Pat$ behaves as the diabatic transition
probability $\Pdt$ in Fig.\ \ref{Fig-diabatic}.
For large $\omega$,
$\Pat$ rises from zero at $-\infty$ to its maximum near the crossing
($\tau=0$) and then decreases to its exponentially small asymptotic
value $\Pai = e^{-\pi\omega^2}$ \cite{note}.
The small-$\omega$ case can be treated in the same manner as in the
diabatic basis, while the large-$\omega$ case requires a more carefull
analysis.

The time evolution of the adiabatic transition probability
before and after the crossing is approximated by the formulas
\bml
\label{Pa}
\bea
\label{Pam}
&&\Pa(\tau<0) \approx \frac {\omega^2} {16 (\tau^2+\omega ^2)^3},\\
\label{Pap}
&&\Pa(\tau>0) \approx e^{-\pi \omega ^2}
	+\left( 1-2e^{-\pi \omega ^2}\right) 
	 \frac{\omega ^2}{16\left( \tau^2+\omega ^2\right)^3}\nonumber\\
&&\qquad +e^{-\pi \omega ^2/2}\sqrt{1-e^{-\pi \omega ^2}}
	 \frac \omega {2\left( \tau^2+\omega ^2\right)^{\case32}}
		\sin \xi(\tau), 
\emla
valid for $\tau^2+\omega^2\gtrsim1$.
The phase $\xi(\tau)$ is given by Eq.\ (\ref{xi}).
Equations (\ref{Pa}), which are new, can be derived from Eq.\ (\ref{Pa-exact})
in the same manner as Eqs.\ (\ref{Pd}) have been derived from
Eq.\ (\ref{Pd-exact}) in \cite{Vitanov97}, but by keeping more terms in
the large-argument-and-large-order asymptotic expansions \cite{Olver58}
of the parabolic cylinder functions in Eq.\ (\ref{Pa-exact}).

\subsection{Jump time}

\subsubsection{Small $\omega$}

For small $\omega$, the transition probability evolution resembles that
in the diabatic basis.
Hence, I define the jump time $\Taj$ in the same way as Eq.\ (\ref{Td-jump-def}),
\be
\label{Ta-jump-def}
\Taj = \frac {\Pai}{\Pa^\prime(0)}.
\ee
By using Eqs.\ (\ref{Pa(inf)}) and (\ref{dPa(0)}) we find that
\be
\label{Ta-jump-small}
\Taj = 2\omega e^{-\pi\omega^2/2} 
	\approx 2\omega, \qquad (\omega^2 \ll 1).
\ee
Hence, for small $\omega$, the jump time in the adiabatic basis is
proportional to $\omega$, as expected.

\subsubsection{Large $\omega$}

For large $\omega$, I define the initial time of the transition as the
time $\Taji<0$ at which $\Pat=\eps \Pai$,
where $\eps$ is a suitably chosen small number.
It follows from Eq.\ (\ref{Pam}) that
\be
\label{Taji}
\Taji  \approx -\omega \sqrt{\left(
	\frac {e^{\pi\omega^2}}{16\eps\omega^4}\right)^{\case13}-1}.
\ee

\begin{figure}[tb]
\vspace*{0mm}
\centerline{\psfig{width=80mm,file=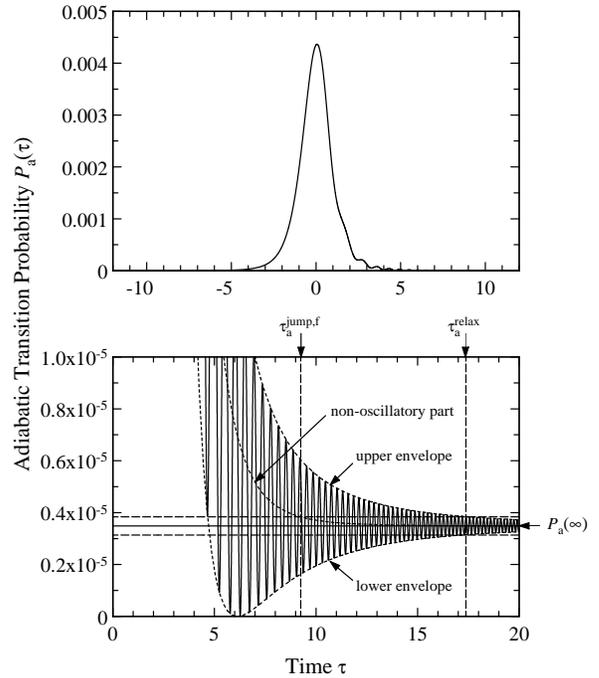}}
\vspace*{-1mm}
\caption{
The time evolution of the adiabatic transition probability $\Pat$
for $\omega=2$.
The upper figure displays the region around the crossing ($\tau=0$).
The lower figure gives an expanded view of the region after the crossing,
which shows the upper and lower envelopes of the oscillations and the
non-oscillatory part of $\Pat$ (short-line dashed curves).
The horizontal solid line depicts the asymptotic value $\Pai$,
whereas the two horizontal dashed lines above and below it show
the values $(1+\eps)\Pai$ and $(1-\eps)\Pai$, respectively, with $\eps=0.1$.
The two vertical dashed lines show the final time of the jump $\Tajf$
[defined by the crossing of the non-oscillatory part of $\Pat$ with
the $(1+\eps)\Pai$-line] and the relaxation time $\Tar$ [defined by
the time when the oscillation amplitude gets smaller than $\eps\Pai$].
}
\label{Fig-omega2}
\end{figure}

To define the final time of the jump $\Tajf>0$, I first remark that
as $\tau$ increases after the crossing, the non-oscillatory part of the
transition probability (\ref{Pap}) approaches the asymptotic value
$\Pai$ from {\it above} [because for $e^{\pi\omega^2}>2$, i.\ e., for
$\omega \gtrsim 0.47$, the second term in Eq.\ (\ref{Pap}) is positive].
Hence, I define $\Tajf$ as the time at which the non-oscillatory part of
$\Pat$ is equal to $(1+\eps)\Pai$.
An illustration of this definition is shown in Fig.\ \ref{Fig-omega2}.
A simple calculation gives
\be
\label{Tajf}
\Tajf \approx \omega \sqrt{\left(\frac {e^{\pi\omega^2}-2}
		{16\eps\omega^4} \right)^{\case13}-1}.
\ee
The total jump time is
\be
\label{Ta-jump-large}
\Taj = \Tajf - \Taji.
\ee
For $\omega^2 \gg 1$, we have
\be
\label{Taj-approx}
\Taj \approx \left(\frac 4 \eps\right)^{\case16}
		\omega^{\case13}e^{\pi\omega^2/6},\quad (\omega^2 \gg 1).
\ee
Thus, for large $\omega$, the transition time in the adiabatic basis is
not proportional to $\omega$, but it rather increases exponentially.
This behavior can be explained by the fact that for large $\omega$,
$\Pai$ ($= e^{-\pi\omega^2}$) is exponentially small and then
the population changes in the slowly vanishing wings of the non-adiabatic
coupling $\thi^{\prime}(\tau)$ [see Eq.\ (\ref{Omega0-dtheta})]
are non-negligible compared to $\Pai$.

\subsection{Relaxation time}

As Eq.\ (\ref{Pap}) shows, at large positive times the amplitude of the
oscillations, that appear in $\Pat$ after the crossing,
vanishes as $\tau^{-3}$.
The relaxation time $\Tar$ is defined in the same way as in the diabatic
basis -- as the time it takes to damp the oscillation amplitude
to the (small) value $\eps \Pai$.
By using Eq.\ (\ref{Pap}), we find
\be
\label{Ta-relax}
\Tar \approx \omega \sqrt{\left(
	 \frac{e^{\pi \omega ^2}-1}{4\eps ^2\omega ^4}\right)^{\case13}-1}.
\ee
For small and large $\omega $, this equation reduces to
\bmla
\label{Tar-small}
&&\Tar \approx \left( \frac{\pi}{4\eps^2}\right)^{\case16}\omega^{\case23},
	\qquad (\omega^2 \ll 1), \\
\label{Tar-large}
&&\Tar \approx \left(\frac 1{2\eps}\right)^{\case13}\omega^{\case13}
	e^{\pi \omega^2/6}, \qquad (\omega^2 \gg 1). 
\emla

A comparison of Eqs.\ (\ref{Td-relax}) and (\ref{Tar-small}) shows that
for small $\omega$, $\Tdr \gg \Tar$.
This is explained by the fact that the oscillation amplitude of
$\Pat$ vanishes as $\tau^{-3}$, i.\ e., much faster than that
of $\Pdt$ which vanishes as $\tau^{-1}$.
In contrast, for large $\omega$, we have $\Tar \gg \Tdr \approx 0$,
which follows from the fact that the reference value in the diabatic basis
is $\Pdi = 1-e^{-\pi\omega^2} \approx 1$, while the reference value
in the adiabatic basis is $\Pai = e^{-\pi\omega^2} \ll 1$.

\begin{figure}[tb]
\vspace*{0mm}
\centerline{\psfig{width=70mm,file=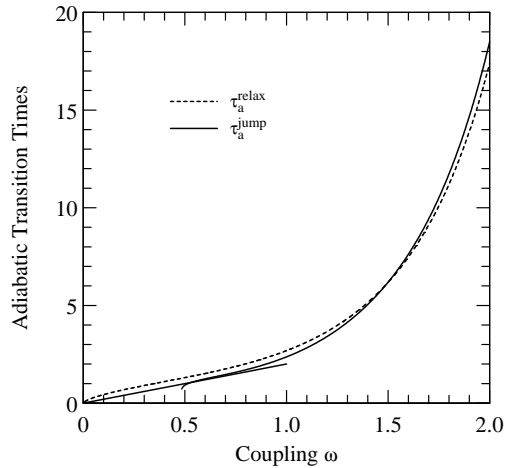}}
\vspace*{-1mm}
\caption{
The jump and relaxation times of the adiabatic transition probability,
plotted against the dimensionless coupling $\omega \equiv \Omega/\beta$.
The small-$\omega$ jump time (\protect\ref{Ta-jump-small})
is shown by the solid line in the range $0 \leqq \omega \leqq 1$,
whereas the large-$\omega$ jump time (\protect\ref{Ta-jump-large})
is shown by the solid curve for $\omega \geqq 0.5$.
Note that the small-$\omega$ and large-$\omega$ formulas
agree well for $0.5<\omega<1$.
The relaxation time (\protect\ref{Ta-relax}) is plotted by a dashed curve.
}
\label{Fig-Ta}
\end{figure}

The adiabatic jump time $\Taj$ and the relaxation time $\Tar$
(for $\eps=0.1$) are displayed in Fig.\ \ref{Fig-Ta}.
Note that, as follows from Eqs.\ (\ref{Taj-approx}) and (\ref{Tar-large}),
the ratio between the jump and relaxation times at large $\omega$
is constant, $\Taj/\Tar \approx (16\eps)^{\case16}$, i.\ e.,
they are almost equal for $\eps=0.1$.
In the evolution picture (Fig.\ \ref{Fig-omega2}), however,
the relaxation ends later than the jump, because the jump time
is calculated from $\tau=\Taji<0$, while the relaxation time
is calculated from the crossing ($\tau=0$).

\section{Summary of the results and conclusions}

\label{Sec-conclusion}

In the present paper, I have calculated various transition times in the
Landau-Zener model.
I have emphasized the differences between the transition times in the diabatic
and adiabatic bases, and between the jump time and the relaxation time.
The {\it jump time} is the time for which the transition probability
rises to the region of its asymptotic value $P(\infty)$.
The  {\it relaxation time} is the time for which the amplitude of the
oscillations, which appear in the transition probability after the crossing,
is damped below the (small) value $\eps P(\infty),\ (\eps \ll 1)$.
These transition times have been calculated by using the exact values of
the transition probabilities and their derivatives at the crossing point
as well as approximations to the transition probabilities evolutions
derived in \cite{Vitanov97} and here.

The results for the jump time in the diabatic basis $\Tdj$ confirm the
scaling properties found by Mullen {\it et al} \cite{Mullen89} in the
limits of small and large coupling $\omega$, i.\ e., for large $\omega $,
$\Tdj$ is proportional to $\omega $, whereas for small $\omega $, $\Tdj$
is constant.
The jump time in the adiabatic basis $\Taj$ has a rather
different dependence on $\omega$.
The seemingly obvious conclusion that $\Taj$ should be proportional to
$\omega$ (because the non-adiabatic coupling is a Lorentzian function of
time with a width of $2\omega$ and the energies of the two adiabatic
states have an avoided crossing with the same duration) turns out to be
correct for small $\omega$ only, while for large $\omega$, $\Taj$
increases exponentially.

The relaxation times in the two bases, $\Tdr$ and $\Tar$, also show rather
different dependences on $\omega$.
The diabatic relaxation time $\Tdr$ is a decreasing function of $\omega$
and it vanishes above certain $\omega$ ($\approx 1.2$),
whereas the adiabatic relaxation time $\Tar$ is an exponentially increasing
function of $\omega$.

It should be pointed out that the transition times obtained in this work
refer to the {\it transition probabilities} $\Pdt$ and $\Pat$.
These may differ from the transition times for the probabilities of no
transition, $1-\Pdt$ and $1-\Pat$, particularly those times which are
linked to the values of the probabilities at $\tau \rightarrow \infty$.

The present paper has dealt with the transition times in the diabatic
and adiabatic bases only, which are the most frequently used bases in
practical applications of the LZ model.
It has been shown by Berry and Lim \cite{Lim91,Berry90} in their
superadiabatic treatment of quantum evolution that the transition time is
shortest and the oscillations in the corresponding transition probability
are minimal in the optimal superadiabatic basis.

In conclusion, the transition times obtained in this paper provide simple
criteria for estimating the applicability of the Landau-Zener
model to various cases of level crossings and avoided crossings.

\appendix

\section{Numerical integration of the Landau-Zener problem}

\subsection{Diabatic basis}

Since in the LZ model (\ref{LZ}) the coupling does not vanish at infinity
and the detuning approaches infinity very slowly, the numerical integration
of Eq.\ (\ref{SEq}) is not a trivial problem, particularly when high
accuracy is required.
The straightforward way of integrating Eq.\ (\ref{SEq}) is to start at
a certain large negative time and propagate the solution towards $+\infty $.
However, a {\it finite} initial time $\tau_{\rm i}$ generates spurious
oscillations with an amplitude proportional to
$(\tau_{\rm i}^2+\omega^2)^{-\case12}$ \cite{Vitanov97} and one has to take
a very large $\tau_{\rm i}$ in order to achieve a good accuracy in $\Pdt$,
which is very expensive in terms of computation time.
An alternative and much more efficient solution to this problem has been
proposed in \cite{Vitanov97}, which is summarized here for the reader's
convenience.
The transition probability is derived from the equation for
the population inversion $\Wd(\tau)\equiv 2\Pdt-1$ (derived from the optical
Bloch equations \cite{Shore90}),
\be
\label{Wd}
\tau \Wd^{\prime\prime\prime} - \Wd^{\prime\prime}
	+4\tau(\omega^2+\tau^2)\Wd^{\prime} - 4\omega^2\Wd=0, 
\ee
rather than from Eq.\ (\ref{SEq}).
The integration starts at $\tau=0$ and the solution is propagated
towards the desired (positive or negative) time.
The initial conditions are found by identifying the terms
in the Taylor expansion of $\Pdt$ around $\tau=0$
[obtained by using the power series expansions of the parabolic cylinder
functions in Eq.\ (\ref{Pd-exact})] with the derivatives of
$\Pdt$ at $\tau=0$. 
The initial values needed to start a Runge-Kutta algorithm, are
\bml
\bea
&&\Wd(0)=-e^{-\pi \omega^2/2} \\
&&\Wd^{\prime }(0)=2\omega \sqrt{1-e^{-\pi \omega ^2}} \cos \hi \\
&&\Wd^{\prime \prime }(0)=4\omega ^2e^{-\pi \omega^2/2} \\
&&\Wd^{\prime \prime \prime }(0)=4\omega \sqrt{1-e^{-\pi \omega ^2}}
     (\sin \hi -2\omega ^2\cos \hi )
\eea
\eml
where $\hio$ is given by Eq.\ (\ref{chi}).

\subsection{Adiabatic basis}

A similar numerical method, which is new and complements the one described
above for the diabatic basis \cite{Vitanov97}, can be used to obtain the
transition probability $\Pat$ in the adiabatic basis
and it has similar advantages.
It turns out convenient to use the angle
$\thi \equiv \case12\arctan (\omega /\tau)$, rather than the time
$\tau$, as an independent variable.
The equation for the population inversion
$\Wat[\th] = \Wa(\tau) \equiv 2\Pat-1$ has the form
\bea
\label{Wa}
\Wat^{\prime\prime\prime}
	+ 6\cot 2\thi \ \Wat^{\prime\prime}
	+ 4[4\omega^4(\cot^2 2\thi + 1)^3+1] \Wat^{\prime} \nonumber\\
	+ 24\cot 2\thi \ \Wat = 0,
\eea
where a prime now means $d/d\thi$.
The initial values of $\Wat(\thi)$ and its derivatives at
$\thi=\pi/4$ (i.\ e., at $\tau=0$), needed to start a Runge-Kutta
algorithm, are 
\bml
\bea
&&\Wat(\pi /4)=-\sqrt{1-e^{-\pi \omega ^2}}\sin \hi , \\
&&\Wat^{\prime}(\pi /4)=-2e^{-\pi \omega ^2/2}, \\
&&\Wat^{\prime\prime}(\pi /4)
 = 4\sqrt{1-e^{-\pi \omega^2}}\left(\sin\hi -2\omega ^2\cos\hi \right), \\
&&\Wat^{\prime\prime\prime}(\pi /4)=8\left( 4\omega ^4+1\right)
 e^{-\pi \omega ^2/2}. 
\eea
\eml
with $\hio$ given by Eq.\ (\ref{chi}).


\begin{references}
\bibitem{note}  This behavior can be deduced more rigorously from Eqs.\
(\ref{Pa(inf)}) and (\ref{Pa(0)}), which show that for $\omega^2 \gg 1$,
$\Pa(0)/\Pai\approx e^{\pi\omega^2}/(16\omega^4) \gg 1$,
and from Eqs.\ (\ref{dPa(0)}), (\ref{d2Pa(0)}) and (\ref{chi-large}),
which give $\Pa^\prime(0)\approx 0$ and $\Pa^{\prime\prime}(0) < 0$,
that indicates a maximum of $\Pat$ near $\tau=0$.
\end{references}
\end{document}